# A Trust-based Framework for Congestion-aware Energy Efficient Routing in Wireless Multimedia Sensor Networks

Srinjoy Ganguly[1], Arpita Chakraborty[2], and Mrinal Kanti Naskar[1]

[1]Dept. of Electronics and Telecommunication Engineering, Jadavpur University, Kolkata – 700032, India
[2]Dept. of Electronics and Communication Engineering, Techno India, Salt Lake, Kolkata 700091, India

srinjoy_ganguly92@hotmail.com, carpi.technoindia@yahoo.com, mrinalnaskar@yahoo.co.in

*Abstract* – **A new era in wireless sensor network technology has been ushered in through the introduction of multimedia sensor networks, which has a major bottleneck in the form of network congestion. Congestion occurs when resources are in high demand during the active period while the data processing and transmission speeds lag behind the speed of the incoming traffic. This may disrupt normal network operations by buffer overflow, packet loss, increased latency, excessive energy consumption and even worse, a collapse of the entire operation. In this paper we propose a novel Trust Integrated Congestion-aware Energy Efficient Routing algorithm (TCEER) in which the potential of a node is computed using its trust value, congestion status, residual energy, distance from the current packet-forwarding node and the distance from the base station using a Fuzzy Logic Controller. The source node selects the node of highest potential in its one hop radio range for data transmission. Hop by hop data routing from source to base station is obtained which is light-weight as well as energy-efficient. Finally, the merits of the proposed scheme are discussed by comparing it with the existing protocols and the study shows promising improvements in network performance.**

## I. INTRODUCTION

Wireless Multimedia Sensor Networks (WMSNs) are the new, emerging field of Wireless Sensor Networks (WSNs) that contain sensor nodes having low cost CMOS cameras, microphones and other sensor devices for retrieving video and audio streams, still images and scalar sensor data from the physical environment [1]. Similar to WSNs, WMSNs are resource-constrained in terms of battery power, limited memory space, computational capability and communication bandwidth. The densely populated, randomly deployed sensor nodes in WMSNs are heterogeneous in nature and generate large volumes of high bit-rate multimedia data, causing network congestion in the upstream direction leading to buffer overflow, increased latency, packet drops, wastage of energy, deterioration of QoS and lower network throughput. Again, low cost sensor nodes are prone to failure. These faulty nodes, known as malicious nodes create several security threats in the sensor networks. Some security attacks have direct impact on the network congestion. The resulting effect is additional computation and communication overhead and an increase in energy consumption which effectively reduces network lifetime. So, detection and isolation of malicious nodes reduces network congestion and thereby improves the network performance significantly. Trust management is a new idea that has been used to detect the malicious nodes and to select trustworthy data routing from source to BS. The concept of trust is basically borrowed from our human society in which sensor nodes monitor the behavior of their neighbors to establish trust-worthy relationships among themselves [2].

In the proposed model, malicious nodes are detected and isolated so that only trusted nodes can participate in the data routing algorithm. For each trust-worthy node the trust value, the congestion status, the residual energy, the distance from the current data packet-forwarding node and the distance from the BS plays a crucial role in the proposed energy-efficient routing protocol. It deploys a Fuzzy model for extracting two metrics, namely the Trust-Congestion Metric (TCM) and the Energy-Distance Metric (EDM) using trust, CCI, distance and residual energy as the computation metrics. The Node Potential (NP) of the trusted node is using TCM and EDM. Hence, NP represents the potential of the node to become the next neighbor in terms of trust, congestion, residual energy, distance from the current node and distance from the BS. During data packet transmission, the source node selects the node with highest NP towards the BS, within its one hop neighbor. In the next hop, the current source node also selects the node with highest NP in its radio range. In this way, hop by hop data routing is obtained from source to BS. The general architecture of the FLC usually comprises of four components. The Fuzzifier converts the crisp input data to fuzzy sets. The Fuzzy output is obtained from fuzzy inference mechanism by combining fuzzy rules into a mapping routine from input to output of the system. Finally, the Defuzzifier extracts a crisp output value from the output fuzzy set.

The rest of the paper is organized in the following manner. In Section II, a brief description of the related work is presented. The proposed Trust-integrated Congestion-aware Energy Efficient Routing algorithm, TCEER is discussed in Section III. The simulation results, performance evaluation and comparison with peers are included in Section IV. Finally, Section V concludes the paper.

## II. RELATED RESEARCH WORKS

Congestion and security attacks are common phenomena in resource-constrained WSNs, especially for WMSNs, where a large volume of high bit-rate multimedia data needs to be managed by the network. Trust-based congestion aware routing in WSNs is a new research topic and has not been addressed in literature to a great extent. T-LEACH [3] is the improved version of the popularly known data-gathering algorithm, LEACH [4], which minimizes the number of cluster head selections and thus extends the lifetime of the network, compared to that of other similar protocols, but it does not take trust & congestion into account. TRANS [5] describes a suite of routing protocols equipped with trust management. In the FCC protocol [6], Zarei et al. have proposed a fuzzy logic-based trust estimation scheme for congestion control in WSNs. FCCTF protocol [7] is basically a modification of FCC, in which the Threshold Trust value is used for decision making. Our previous work, which shows a major improvement over the FCCTF algorithm, is described in TFCC [8], in which traffic flow from the source to sink is optimized by adaptive data-rate control and data routing takes place by virtue of the Link State Routing Protocol.

## III. THE PROPOSED ALGORITHM: TCEER

In this section we describe the Trust-integrated Congestion-aware Energy Efficient Routing (TCEER) protocol for WMSNs. It is applicable for general purpose WSNs as well. We consider a random deployment of sensor nodes in the sensor field under the condition of free space propagation. The TCEER protocol consists of two phases. Phase I is the initialization phase whereas Phase II is the routing phase. The details of each phase are described below:

*Phase I : Initialization Phase*

The phase I algorithm basically computes four parameters, namely trust, complementary congestion index (CCI), distance metric and energy metric. The malicious nodes are also segregated in this phase on the basis of their trust values.

*Trust Calculation and Segregation of Malicious Nodes*

Trust of a node on another node is evaluated dynamically on the basis of various Trust Metrics (TM) which measures the behavior of the node during previous data transfer through this node, called Direct Trust and the recommendations received from other trusted nodes, known as Indirect Trust. Trust of a node on it's the neighbor nodes within its radio range is calculated by the method described in GMTMSS [9]. Direct Trust of node $N_1$ on node $N_2$ ($DT_{N_1,N_2}$) is calculated from the geometric mean of the various Trust Metrics for different events occurred between $N_1$ and $N_2$. Any number of TM can be considered for trust computation. For *k* TMs it is represented by the formula as given below:

$$DT_{N_1,N_2} = \left(\prod_{i=1->k}(TM_i)\right)^{\frac{1}{k}} \quad (1)$$

Indirect Trust of $N_1$ on $N_2$ ($IT_{N_1,N_2}$) is computed by the geometric mean of various Direct Trusts (DTs), obtained from different neighboring nodes of $N_1$, which is represented as:

$$IT_{N_1,N_2} = \left(\prod_{j=1->l}(DT_j)\right)^{\frac{1}{l}} \quad (2)$$

[$DT_1, DT_2, DT_3 \ldots \ldots DT_l$ are the DTs from *l* number of neighbor nodes of $N_1$]

If $W_D$ and $W_I$ are the weights given to DT and IT respectively, the overall trust of node N1 on N2 ($T_{N1(N2)}$) is the weighted sum of DT and IT which is represented by the formula.

$$T_{N1(N2)} = W_D * DT_{N1(N2)} + W_I * IT_{N1(N2)} \quad (3)$$

A predefined Trust Threshold ($T_{TH}$) value is considered, depending on the application of the sensor nodes. Greater the value of $T_{TH}$, higher is the security of the network. The nodes with the condition $T_{N1(N2)} > T_{TH}$, is termed as trusted nodes whereas, for $T_{N1(N2)} < T_{TH}$, the node is called malicious and are henceforth blocked.

*Computation of the Congestion Metric*

**Student Author** – Srinjoy Ganguly (Fourth Year Undergraduate Student)

The congestion level of the trusted nodes is estimated from the current buffer queue size of the corresponding sensor nodes. Two fixed threshold values, $C_{Th}$ (Min) and $C_{Th}$ (Max) are defined in the range of the queue length. The parameter, called the Complementary Congestion Index (CCI) is calculated which is the function of buffer queue length. It is assumed that each sensor node has only buffer where in stores all the packets that it accepts from its one-hop neighbors. Let the queue size [10] of the $k^{th}$ node be denoted by $Q_s(k)$. Then the complementary congestion index for the $k^{th}$ node (represented by $I'_k$) may be defined mathematically as:

$$I_k' = 1 - I_k$$

where,

$$I_k = \in \quad \text{for, } Q_s(k) < C_{TH}(Min)$$
$$I_k = 1 \quad \text{for, } Q_s(k) > C_{TH}(Max)$$
$$I_k = (1-\in)\left(\frac{Q_s(k) - C_{TH}(\min)}{C_{TH}(\max) - C_{TH}(\min)}\right) + \in \quad \text{for, } Q_s(k) \in [C_{TH}(Min), C_{TH}(M_{AX})]$$

[$\in$ is a small quantity lying between zero and one] (4)

*Calculation of the Energy Metric*

It is considered that the initial energy of all nodes is $E_{initial} = 0.5$ Joules. The effective residual energy ($E_{er}$) of the node is normalized as:

$$E_{er} = \omega * E_{cn} + (1-\omega) * E_{pnn} \quad (5)$$

Here $E_{cn}$ denotes the energy of the current node and $E_{pnn}$ represents the energy of the potential next node. The parameter $\omega$ is the weighing factor, usually set to a value greater than 0.5 so as to give higher priority to the remaining energy of the potential next node. The potential next node means the nearest trusted node from the current node, within its one hop neighbor, in the radio communication range.

*Calculation of the Distance Metric*

Let $d_1$ is the ratio of the distance between current node and potential next node to the radio communication range between the nodes. Similarly, $d_2$ is the distance between potential next node and the BS to the distance between current node and the base station. The parameters $d_1^C$ and $d_2^C$ are called the complementary distances from the current node and the complementary distance from the BS respectively, which are given by the equations:

$$z, \; d_2^C = 1 - d_2 \quad (6)$$

Lower the values of $d_1$ and $d_2$ i.e. higher the values of $d_1^C$ and $d_2^C$, the more desirable is the situation. The distance metric of the potential next node is represented by the equation:

$$Distance \; Metric = \frac{k_1 * d_1^C + k_2 * d_2^C}{k_1 + k_2} \quad (7)$$

Where, $k_1$ and $k_2$ are the weights of $d_1^C$ and $d_2^C$ respectively. In the proposed TCEER algorithm, we have given more importance to the distance from the BS and hence the value of $k_2$ is chosen higher, compared to the value of $k_1$. The potential next node should always be closer to the BS, compared to the current node, to make the routing distance minimum. Hence, the distance between the potential next node and BS is less than the distance between current node and BS. So, the ratio of the distance between potential next node to BS to the Distance between current node and BS is always less than one. It implies that $d_2 < 1$.

*Phase II: Route Determination Phase*
*Computation of Node Potential and Data Packet Routing*

In this phase, a Fuzzy Logic Controller (FLC) is used for the computation of the Trust Congestion Metric (TCM) and the Energy Distance Metric (EDM) of the trusted nodes. The configuration of the FLC that is used in the TCEER scheme is shown in Fig. 1. It consists of a Fuzzifier-1/ Defuzzifier-1/ Rule Base-1/ Inference Mechanism-1 and a Fuzzifier-2/ Defuzzifier-2/ Rule Base-2/ Inference Mechanism-2, respectively. The inputs to the FLC are the parameters that have been computed in phase I. Node Potential (NP) of a trusted node is calculated as:

**Student Author –** Srinjoy Ganguly (Fourth Year Undergraduate Student)

$$\text{Node Potential} = \frac{\alpha * EDM + \beta * TCM}{\alpha + \beta} \qquad (8)$$

Where, α and β are the weights assigned to EDM and TCM, respectively on the basis of the application of the sensor networks, in which summation of α and β is equal to unity.

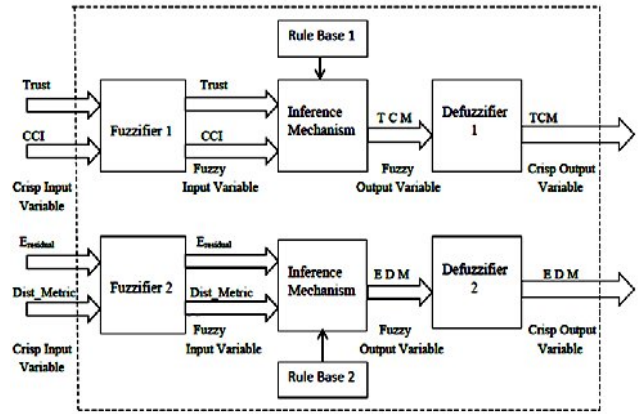

**Fig.1. Schematic diagram of the FLC used in TCEER**

The data routing in TCEER is done on the basis of the node potential (NP) within the one-hop neighbors. Whenever the onus upon the current node is to forward a data packet, it will select the next intermediate node having highest NP in its set of one-hop neighbor. In this way, packet transmission will continue until the receiving node becomes the BS.

## IV. EXPERIMENTAL RESULTS

In this section, the merits of the proposed TCEER scheme have been investigated through extensive MATLAB simulations. We have considered an arbitrary network, comprising of 50 multimedia sensor nodes deployed randomly into a field of dimension 200 m * 200 m. In the simulation experiments, we have taken three TMs, namely data packets forwarded, packet addresses modified and remaining energy of the nodes. Initially, we have considered that the nodes are all trusted nodes. The trust value of the node is updated periodically after Δt seconds. The Trust Threshold ($T_{Th}$) value is taken as 0.5 whereas minimum and maximum trust values are 0 and 1 respectively. The values of the constant parameters are chosen as: $\alpha = 0.3$, $\beta = 0.7$, $\omega = 0.2$, $K_1 = 2$ and $K_2 = 3$. Fig. 2 shows the graphical view of the packet routing, when 20 data packets are transmitted from the source node, marked as 24, towards the BS. It is found that the packets have taken different routes to reach the BS, at different time, depending on the value of the NP which is modified dynamically.

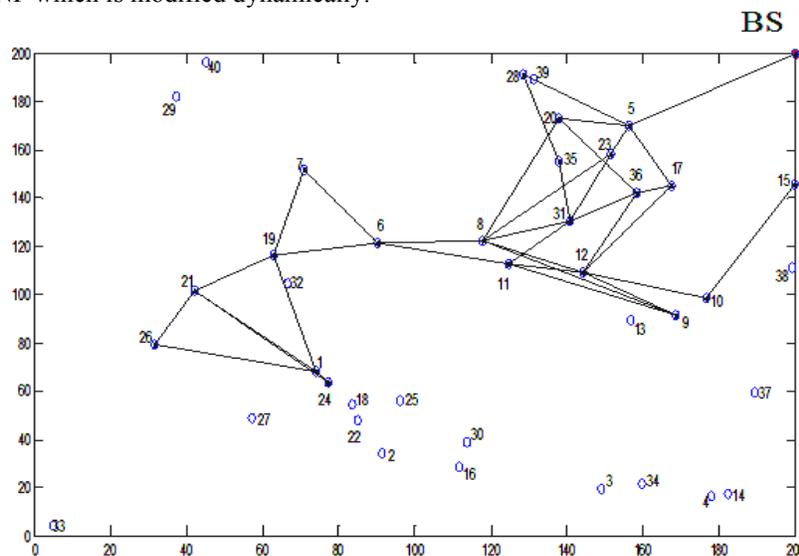

**Fig.2. Routing of 20 packets from source node 24**

**Student Author** – Srinjoy Ganguly (Fourth Year Undergraduate Student)

The proposed TCEER protocol is compared with the existing algorithms, such as T-LEACH [3], TRANS [5] and TFCC [8]. The number of rounds verses percentage of dead nodes for the above mentioned protocols have been shown graphically in Fig. 3. The simulation and experimental results indicate that the TCEER protocol provides higher network lifetime compared to other similar protocols and thereby outperforms its peers.

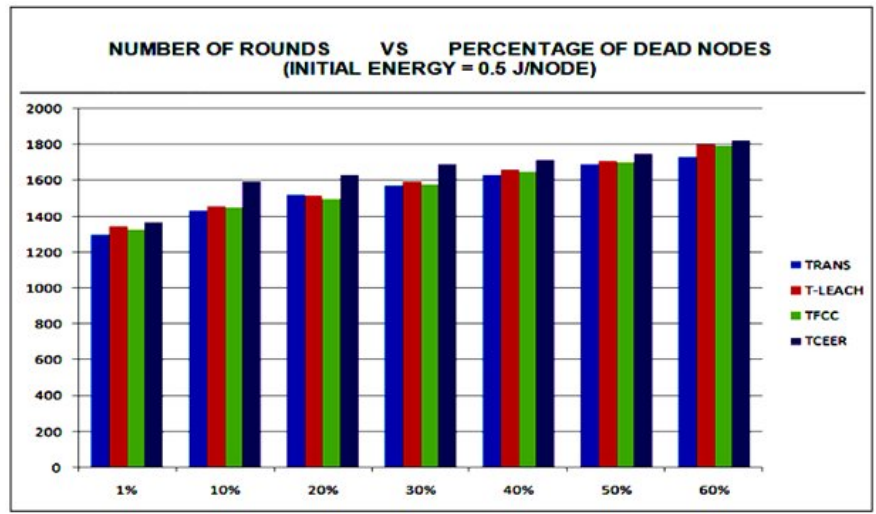

**Fig. 3 Performance Analysis of Various Protocols with Initial Energy of 0.5 Joule per Node**

## V. CONCLUSION

In this paper we have presented a novel trust-based congestion aware routing protocol using a Fuzzy Logic Controller for WMSNs, which is also applicable for large WSNs. A significant improvement in network lifetime has been achieved by taking trust and congestion into account. Thus, the optimum route for data packet transfer is dynamically selected on the basis of Node Potential (NP). The experimental results show that the proposed algorithm provides a higher number of rounds, indicating higher network lifetime compared to the other protocols which it has been compared with, due to the efficiency with which it detects routes with higher potential. However, the TCEER protocol has been tested only on a small network. In the future, we would like to test its impact on larger networks comprising of large number of heterogeneous multimedia sensor nodes. We also desire to test its hardware implementation with Iris motes using TinyOS under various conditions.

**Student Author –** Srinjoy Ganguly (Fourth Year Undergraduate Student)